# CLASSIFICATION OF NASOPHARYNGEAL CASES USING DENSENET DEEP LEARNING ARCHITECTURE

W. S. H. M. W. AHMAD[1,*], M. F. A. FAUZI[1], M. K. ABDULLAHI[1], JENNY
T. H. LEE[2], N. S. A. BASRY[2], A YAHAYA[3], A. M. ISMAIL[2], A. ADAM[4],
ELAINE W. L. CHAN[5], F. S. ABAS[6]

[1]Faculty of Engineering, Multimedia University, 63100 Cyberjaya, Malaysia
[2]Department of Pathology, Sarawak General Hospital, 93586 Kuching, Sarawak, Malaysia
[3]Department of Pathology, Faculty of Medicine, Universiti Kebangsaan Malaysia, Kuala Lumpur, 56000, Malaysia
[4]Center for Artificial Intelligence Technology, Faculty of Information Science and Technology, Universiti Kebangsaan Malaysia, 43600 Bangi, Malaysia
[5]Fusionex AI Lab, International Medical University, 57000 Kuala Lumpur, Malaysia
[6]Faculty of Engineering and Technology, Multimedia University, 75450 Ayer Keroh, Melaka, Malaysia
*Corresponding Author: wan.sitihalimatul@mmu.edu.my / wshmunirah@gmail.com

## Abstract

Nasopharyngeal carcinoma (NPC) is one of the understudied yet deadliest cancers in South East Asia. In Malaysia, the prevalence is identified mainly in Sarawak, among the ethnic of Bidayuh. NPC is often late-diagnosed because it is asymptomatic at the early stage. There are several tissue representations from the nasopharynx biopsy, such as nasopharyngeal inflammation (NPI), lymphoid hyperplasia (LHP), nasopharyngeal carcinoma (NPC) and normal tissue. This paper is our first initiative to identify the difference between NPC, NPI and normal cases. Seven whole slide images (WSIs) with gigapixel resolutions from seven different patients and two hospitals were experimented with using two test setups, consisting of a different set of images. The tissue regions are patched into smaller blocks and classified using DenseNet architecture with 21 dense layers. Two tests are carried out, each for proof of concept (Test 1) and real-test scenario (Test 2). The accuracy achieved for NPC class is 94.8% for Test 1 and 67.0% for Test 2.

## 1. Introduction

Nasopharyngeal carcinoma (NPC) is a type of head and neck cancer. It starts in the nasopharynx, the upper part of the throat behind the nose and near the base of the skull. NPC is associated with Epstein-Barr virus (EBV), but EBV alone is not a sufficient cause for this malignancy [1]. Environmental exposures and genetic factors likely play a role in the pathogenesis of this rare case and make it hard to study [2]. It is endemic to Eastern and South-Eastern Asia, where 77% of global cases were diagnosed in these regions in 2020, with a total of 133,354 new cases and 80,008 deaths. Because of its late symptoms and anatomical location, it makes it difficult to be detected in the early stages [3]. NPC is asymptomatic at the early stage with common symptoms such as swollen lymph node, blood mixed with saliva (epistaxis), nasal congestion or ringing in the ears, hearing loss, frequent ear infections and sore throat [1]. A sample of nasopharyngeal tissue is biopsied for examination to confirm the case (refer Fig. 1). The pathologist will examine the processed tissue which is stained with hematoxylin and eosin (H&E) and mounted on a glass slide under the microscope for diagnosis. H&E is the most commonly used stain and it does not capture cancer receptors like immunohistochemistry stains, but only cells morphology, and requires expert interpretation to determine the malignancy in cells. Since the case is very rare, it takes time to build the level of confidence for an accurate diagnosis. With digital pathology and computer-assisted method, provided good training samples and ground truth, it will perform consistently and can serve as a second opinion to the pathologists.

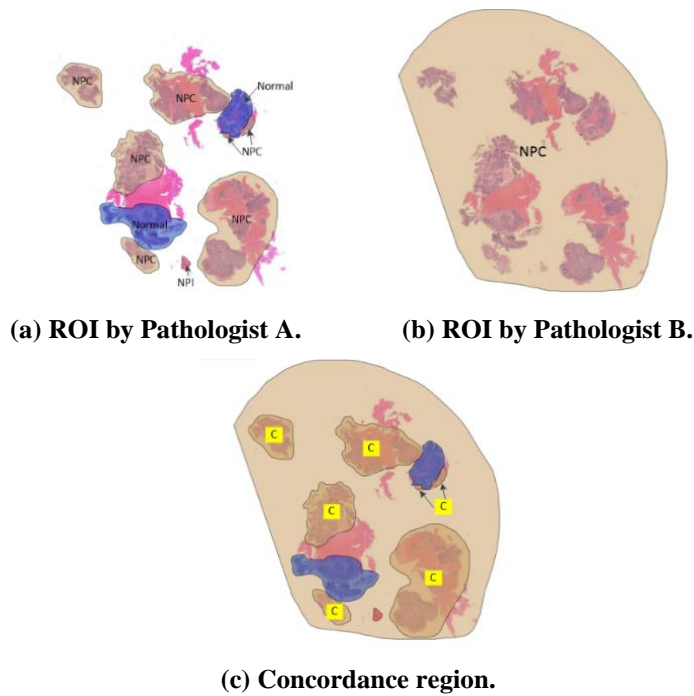

(a) ROI by Pathologist A.          (b) ROI by Pathologist B.

(c) Concordance region.

**Fig. 1. Example of ROI annotations on the WSI by the pathologists. Regions marked as C in yellow boxes are in concordance with the two pathologists.**

A systematic review on segmentation, classification and prediction composition related to NPC has been done by [4] for papers published between 1970 to 2016, and the data mainly consists of magnetic resonance (MR) or computed tomographic (CT) imaging modality of patient's head. A more recent review was carried out by Ng et al. (2022) focusing on artificial intelligence for NPC management from 2006 to 2021. From the selected 60 articles, 1 of them was using microscopic images [5], 3 of them were analysing pathological whole slide images (WSIs) [6,7,8] and another 1 was having multi-cohort study with WSI, MRI and clinicopathological data [9]. All previous work on WSIs was published in 2020, originated from Malaysia, Taiwan and China, with the aim of classification, assessing survival risk in NPC patients, and prognosis. Various methods have been tested, for example K-means clustering and artificial neural network (ANN) [5], deep convolutional neural network (CNN) on patch- and slide-level [6], Inception-v3 [7], DeepSurv as the state of the art survival method for personalized treatment choices [8,9] and Resnet-18 [9].

Based on the literature, the analysis of NPC cases using WSI has four types of diagnoses: normal, NPC, nasopharyngeal inflammation (NPI) and lymphoid hyperplasia (LHP). Since NPC is rare compared to other cancers, no public dataset of pathological images is available to date for algorithm benchmarking. In this study, we have our private dataset from collaborating hospitals, with the aim of classifying NPC cases among other classes (normal, NPI and LHP) from WSIs. However during our inital data acquisition, there is no concordance on the LHP case among the pathologists, hence it is not included.

The novelty of this paper lies on the type of image data used with the deep learning classifier (DenseNet-21) where this architecture has yet been tested in the literature for pathological images. NPC data itself has its own novelty because it is very rare with less than 1 case in 100,000 in the United States [10], and highly prevalent only in the South East Asia. Access to this case is limited where not many hospitals has suffice data for analysis. This work will contribute to address the classification of the different characteristics of nasopharynx biopsies, particularly cases in the prevalent state located in East Malaysia (Sarawak).

## 2. Classification of Nasopharyngeal Cases

The overall process flow is illustrated in Fig. 2. There are three main stages in the workflow: i) ground truth preparation; ii) image dataset establishment; and iii) model training and testing.

### 2.1. Ground truth preparation

This preliminary image database is obtained from two hospitals: 3 whole slide images (WSIs) from Sarawak General Hospital (SGH) and 4 WSIs from Hospital Kuala Lumpur (HKL), as tabulated in Table 1. 3 pathologists were involved in the ground truth annotations: Pathologist A annotated 7 WSIs, Pathologist B annotated 4 WSIs and Pathologist C annotated 3 WSIs. They were asked to identify any of the 3 classes in the WSIs using our Cytomine web-UI collaborative platform [11]. The classes are normal, nasopharyngeal inflammation (NPI) and nasopharyngeal carcinoma (NPC). After the annotations, we found that at least 2 pathologists agreed on the diagnostics for each WSI: 1 with Normal, 1 with NPI and 5 with NPC. The example of WSI with pathologist A and J annotations, as well as their concordance (marked as C in yellow box), are shown in Fig. 1.

The ground truth annotations by pathologists vary in terms of size and precision. This can be seen from Fig. 1 (a) and (b) where one pathologist considered the whole tissue stain as NPC, and the other pathologists carefully delineated parts of the tissue region as different diagnosis. The pathologists are from different hospitals and the annotated regions are marked based on their experience and level of confidence.

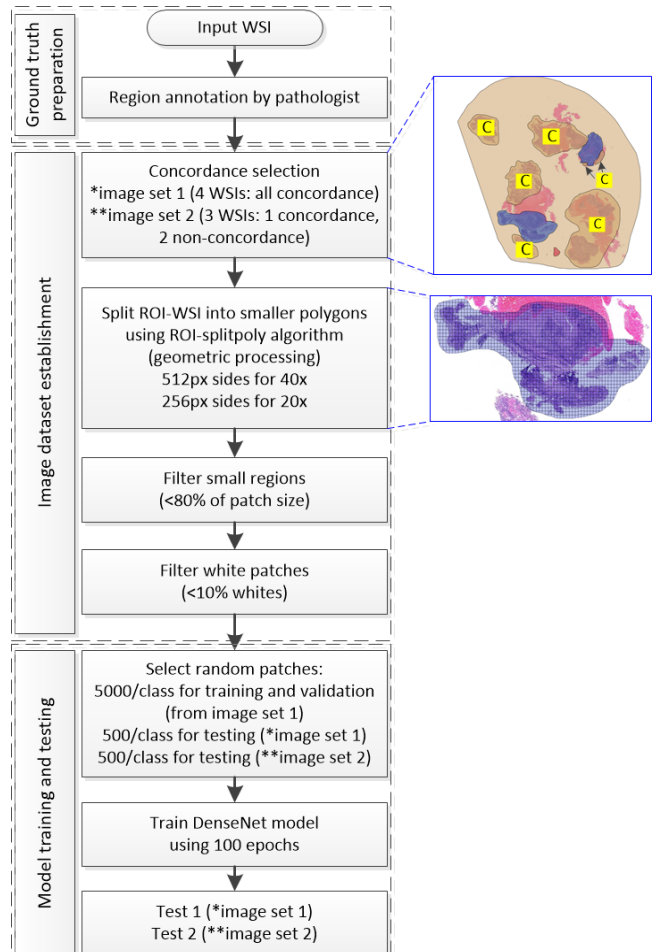

**Fig. 2. Flowchart for the overall process of this preliminary experiment.**

## 2.2. Image dataset establishment

The WSIs acquired from the two collaborating hospitals has different setting, resolution and magnification. For WSIs from SGH, it was scanned using 3DHistech Pannoramic SCAN 150 at 20x magnification, resulting a dimension of approximately 125,000 pixels width and 295,000 pixels height. While for WSIs from HKL are from 3DHistech Pannoramic MIDI scanner at 40x magnification, with the dimension of 250,000 pixels width and 572,000 pixels height. Since the WSIs from HKL are twice larger than SGH with higher magnification, the region of interest (ROI) will be resized into half of the ROIs from SGH for standardization.

For this experiment, the pathologists have done their best to annotate regions according to their diagnosis, hence the patches extracted from the regions are labelled as per provided diagnosis, as shown in Fig. 1 (c). From pathologists' annotations, we established two image sets for experimental purposes. Image set 1 consists of WSI number 1 to 4, with all concordance regions of the two pathologists. Image set 2 is of WSI number 5 to 7, but regions for image 5 and 7 are non-concordance regions (Normal and NPI). We set two different experiments from the two image sets due to the limitation in region concordance. The annotated regions from both image set 1 and image set 2 are then processed with the following steps:

1. Divide large regions into smaller patches:
   a. Condition 1:
      If the large region (ROI-WSI) width ($X_{ROI-WSI}$) or height ($Y_{ROI-WSI}$) is greater than 10,000 pixels, it will be divided into a maximum of 4096 pixels polygons-side ($P_{X|Y}$) prior to smaller patches. This step is to accelerate the division process with a larger chunk of patches.
      Condition 2:
   b. If the ROI-WSI is smaller than 10,000 pixels in width or height, it will be divided into a maximum of 512 pixels or 256 pixels polygons-side for WSI with 40x or 20x magnification ($M$) respectively.

   These conditions can be summarized in the following statement:

   if $X_{ROI-WSI} > 10,000$ or $Y_{ROI-WSI} > 10,000$ :

   $P_{X|Y} = 4,096;$

   else

   $$P_{X|Y} = \begin{cases} 256, & if \quad M = 20 \\ 512, & if \quad M = 40 \end{cases}$$

2. Some of the patches are with exact square, and some with polygon sides, depending on the cutting. To eliminate small patches, only the ones with an area of at least 80% of 512x512 pixels or 256x256 pixels are considered.

3. The patches are then filtered to get the most of the tissue region. Those containing more than 10% white area are removed using a quick 16-bin grey level histogram. If the 16-th bin (white bin) has more than 10% of the patch total pixels, it will be discarded.

4. Patches from 40x magnification WSIs will be resized into half of the original size to standardize with 256x256 pixels patches.

After the division and filtering processes, the resulting number of patches for image set 1 is 37,997 (18,100 NPC, 7,214 Normal and 9,683 NPI) and image set 2 is 6,903 (4,214 NPC, 898 Normal and 1,788 NPI), as summarized in Table 1. These patches will be randomly chosen for each class for training and testing of the deep learning model, as elaborated in the following section.

**Table 1. Summary of image database.**

| Image Set | WSI No. | Concordance region-class annotation (Pathologist) | Non-concordance region-class annotation (Pathologist) | Considered region | Magni-fication | Total patches |
|---|---|---|---|---|---|---|
| Image set 1 | 1 (SGH) | NPC (C), NPC (A) | - | NPC (C)* | 20x | 18,100 |
| | 2 (SGH) | NPC (C), NPC (A) | - | NPC (C)* | | |
| | 3 (SGH) | Normal (C), Normal (A) | - | Normal (C)* | 20x | 7,214 |
| | 4 (HKL) | NPI (B), NPI (A) | NPC (A) | NPI (A)* | 40x | 9,683 |
| Image set 2 | 5 (HKL) | NPC (B), NPC (A) | NPI (A) | NPI (A)** | 40x | 1,788 |
| | 6 (HKL) | NPC (B), NPC (A) | - | NPC (A)** | 40x | 4,214 |
| | 7 (HKL) | NPC (B), NPC (A) | Normal (A), NPI (A) | Normal (A)** | 40x | 898 |

## 2.3. Model training and testing

Regions from image set 1, with single asterisk (*) in Table 1, are considered for model training and validation. A total of 15,000 patches (5,000 patches per class) from the 37,997 patches are randomly selected for training, with 90-10 split for training (90%: 13,500) and validation (10%: 1,500). For example, from 18,100 patches from NPC class, 5,000 patches are randomly selected for the training. DenseNet architecture with 21 dense layers with (2,2,2,2) dense block configuration is used in this preliminary experiment using PyTorch framework. The model is not as deep as DenseNet-201 with 201 layers, but gave better performance with smaller model size and shorter training time. The input image is resized to the shape of 224 by 224 and the model is trained from scratch without pretrained weight with data augmentations from Albumentations [12] using 100 epochs and 64 batch-size. Default parameters of Adam optimizer are used with learning rate of 0.001 and the loss function is calculated using categorical cross-entropy for classification of the three classes.

For model testing, two experiments are carried out, called Test 1 and Test 2. Test 1 is for proof of concept and Test 2 is for real-test, where both are having different test sets. Because of this, we set up two different experiments. Details of each test set are as below and a summary with some examples of the patches for Test 1 and Test 2 are shown in Table 2.

1.  Test 1 (proof of concept): 500 patches per class are randomly selected from image set 1, but not within the 5000 training set patches. Note that this set came from two different sources and magnifications (Normal and NPC from SGH at 20x, and NPI from HKL at 40x). Patches from both sources are standardized to a maximum of 256x256 pixels.

Test 2 (real-test): Similar to Test 1, 500 patches per class are randomly selected from image set 2 (with double asterisk (**) in Table 1 containing the non-concordance regions (for Normal and NPI). All images from this set are sourced from HKL at 40x magnification. The patches are originally at 512 pixels maximum sides but were resized to 256 pixels.

**Table 2. Summary of Image set 1 and Image set 2 patches for Train, Test 1 and Test 2.**

| Image set 1: Train (5000 patches per class) Test 1 (500 patches per class) | | |
|---|---|---|
| **Diagnosis (Source: magnification)** | Normal (SGH: 20x) | NPI (HKL: 40x) | NPC (SGH: 20x) |
| **Patch size** | 256px | 512px resized to 256px | 256px |
| **Concordance of at least 2 pathologists** | Yes | Yes | Yes |
| **Sample images** | 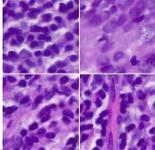 | 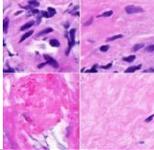 | 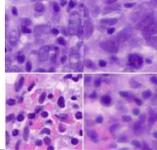 |
| **Image set 2: Test 2 (500 patches per class)** | | |
| **Diagnosis (Source: magnification)** | Normal (HKL: 40x) | NPI (HKL: 40x) | NPC (HKL: 40x) |
| **Patch size** | 512px resized to 256px | 512px resized to 256px | 512px resized to 256px |
| **Concordance of at least 2 pathologists** | **No** | **No** | Yes |
| **Sample images** | 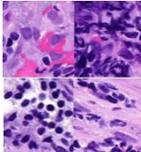 | 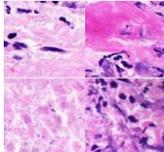 | 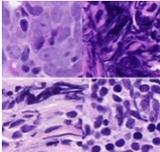 |

## 3. Results and Discussions

### 1. Training and validation

Model training loss plot for 100 epochs is shown in Fig. 3 (a), while validation loss and accuracy plots in (b) and (c). Comparing the two loss plots, there are some fluctuations in the validation set, specifically at epoch 23, 29, 36, 44, 65, 67, 69, 77 and 87, due to high variability of tissue appearance in the WSIs and no clear distinct of appearance between the classes, especially Normal and NPC. The lowest validation accuracy is 63.4% at epoch 65 and highest is 99% at epoch 89. Taking the best model to evaluate the test set, the results for Test 1 and Test 2 are elaborated in the following subsections.

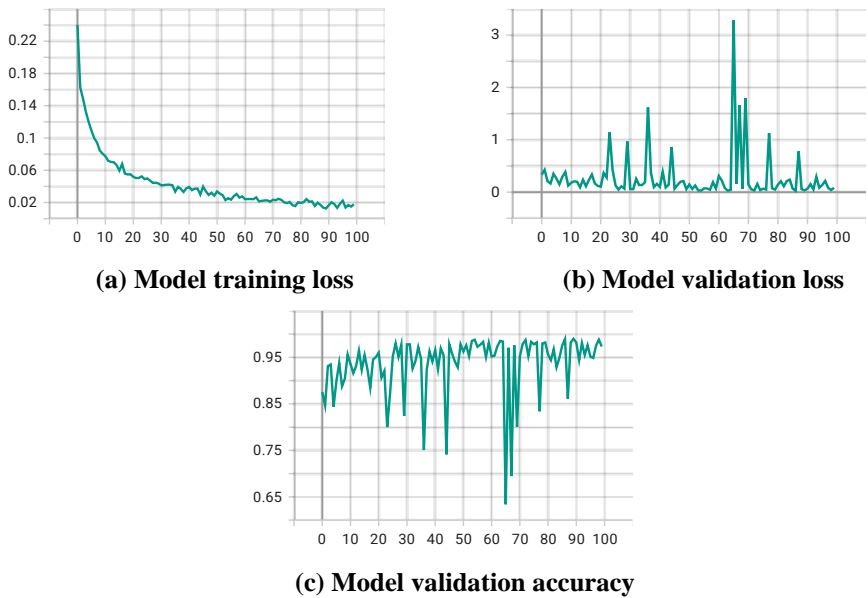

**(a) Model training loss**          **(b) Model validation loss**

**(c) Model validation accuracy**

**Fig. 3. Loss and accuracy plots for model training and validation.**

## 2. Test 1: proof of concept

Fig. 4 shows the confusion matrix table, where each column of the matrix represents the predicted class, and the rows are the true class. The confusion matrix of Test 1 prediction results using DenseNet model with 21 layers is shown in Fig. 4 (a). Out of 500 images in the Normal class, 467 is correctly predicted which correspond to an accuracy of 93.4%, and the remaining 30 and 3 images are wrongly predicted as NPC and NPI respectively. NPI class gave a perfect 100% accuracy, and NPC class has 94.8% correct predictions with 4.4% (22 images) wrongly predicted as NPC. The accuracies for Test 1 are very high because there is not much variety in the training and testing data. The model was trained using patches from image set 1, which also the source of Test 1 patches (same WSIs, different patches).

## 3. Test 2: real-test

For Test 2, we can see the robustness of the minimally trained DenseNet model in the confusion matrix (Fig. 4 (b)). For NPI class, since both are from the same source (HKL-40x), they exhibit similar characteristics resulting in high accuracy (99.2%). NPC class achieved 67% accuracy with 335 correctly predicted patches out of 500 despite the model was trained using patches from different source (SGH). For Normal class, the accuracy is very low with only 9 correctly predicted patches out of 500 (1.8%) with false positive biased to NPI class (0.832%). Looking into the patches visually, some are similar to either NPI or NPC. It is also important to note that this class is non-concordance of the two-assigned pathologists (Fig. 1 (a) and (b)) where one pathologist marked the whole regions as NPC and another pathologist delineated each region as NPC, Normal and NPI. Another reason to the NPI-bias is that NPI patches in the training set are from the same source with the Test 2 image set (HKL-40x), having visually similar features especially the colour and sharpness, as can be seen in Table 2.

It is also understandable that WSI from only 7 patients is insufficient to train a good model, but this is our first initiative for this work, and more elaborate study including image normalization, augmentation and class weighting can be performed with more training samples and ground truth. Our upcoming work will include more WSIs from both SGH and HKL in the training set when we have the ground truth ready from the collaborating pathologists.

| True class | | Predicted class | | |
|---|---|---|---|---|
| | | Normal | NPI | NPC |
| | Normal | 467 (0.934) | 3 (0.006) | 30 (0.06) |
| | NPI | 0 (0.0) | 500 (1.0) | 0 (0.0) |
| | NPC | 22 (0.044) | 4 (0.008) | 474 (0.948) |

**(a) Test 1 (proof of concept).**

| True class | | Predicted class | | |
|---|---|---|---|---|
| | | Normal | NPI | NPC |
| | Normal | 9 (0.018) | 416 (0.832) | 75 (0.15) |
| | NPI | 0 (0.0) | 496 (0.992) | 4 (0.008) |
| | NPC | 2 (0.004) | 163 (0.326) | 335 (0.67) |

**(b) Test 2 (real-test).**

**Fig. 4. Confusion matrix for prediction results using DenseNet model.**

## 4. Conclusions

This preliminary experiment shows that some initial results that can be analyzed from the image dataset obtained from different sources with different machines and configurations. Two experiments have been carried out: Test 1 for proof of concept, and Test 2 as a real-test scenario, with NPC prediction accuracies of 94.8% and 67.0% accordingly. It is worth to note that for Test 2 set, NPI and Normal cases are of non-concordance regions (only one pathologist provide the diagnoses), hence the test result might be predicting the actual diagnosis. We can see that the model is able to correctly classify 67% of NPC cases (concordance of two pathologists) despite it is trained using different magnification level, color and sharpness. The main challenge here is the need for accurate and consistent identification by pathologists for this area in comparison with human variability of interpretation. In this paper, the untuned hyperparameters are used to see how the default model performed. Further experiments will be done to include these, with more cases for each diagnosis. There are also many other potential experimental works can be done provided more time, resources, and collaborative assistance by the pathologists.

| Nomenclatures | |
|---|---|
| $X_{ROI\text{-}WSI}$ | Large region (ROI-WSI) width |
| $Y_{ROI\text{-}WSI}$ | Large region (ROI-WSI) height |
| $P_{X/Y}$ | Polygon sides |
| $M$ | Magnification |

| Abbreviations | |
|---|---|
| H&E | Hematoxilyn and Eosin |
| HKL | Hospital Kuala Lumpur |
| LHP | Lymphoid Hyperplasia |
| NPC | Nasopharyngeal Carcinoma |
| NPI | Nasopharyngeal Inflammation |
| ROI | Region of interest |
| SGH | Sarawak General Hospital |
| WSI | Whole slide image |

## Acknowledgments

We would like to thank our collaborators for providing the image dataset and their ground truth for evaluation.

## Funding Acknowledgments

This work is supported by the Ministry of Higher Education (MOHE) Malaysia through the Fundamental Research Grant Scheme (FRGS/1/2020/ICT02/MMU/02/10) and IsDB-STI Transform Fund.